\documentclass[final,11pt,3p,sort&compress,twocolumn]{elsarticle} 
\makeatletter
\def\ps@pprintTitle{%
 \let\@oddhead\@empty
 \let\@evenhead\@empty
 \def\@oddfoot{\reset@font\hfil\thepage\hfil}%
 \let\@evenfoot\@oddfoot}

\biboptions{super} 

\usepackage[latin2]{inputenc}
\usepackage{amssymb}
\usepackage{amsthm}
\usepackage{amsmath}
\usepackage[version=3]{mhchem} 
\usepackage[T1]{fontenc}       

\usepackage{graphicx,cancel}
\usepackage{ulem}
\usepackage{color}
\usepackage{fixltx2e} 
\usepackage{balance} 
\usepackage{float}	
\usepackage{dblfloatfix} 
\usepackage{booktabs}
\usepackage[size=small, labelfont=bf]{caption} 
\usepackage{hyperref}
\hypersetup{
    colorlinks=true, 
    linkcolor=blue,  
    citecolor=blue,  
}
    
\renewcommand{\_}[1]{{}_{\mathrm{#1}}}

\begin{document}
\begin{frontmatter}

\title{\LARGE{\textbf{Self-Metalation of Phthalocyanine Molecules with Silver Surface Atoms by Adsorption on Ag(110)}}}

\author[a]{\large{Lars~Smykalla}}
\author[a]{Pavel~Shukrynau}
\author[b]{Dietrich R. T. Zahn}
\author[a]{Michael Hietschold}
\address[a]{Technische Universit\"at Chemnitz, Solid Surfaces Analysis Group, D-09107 Chemnitz, Germany}
\address[b]{Technische Universit\"at Chemnitz, Semiconductor Physics, D-09107 Chemnitz, Germany}

\begin{abstract}

We report that metal-free phthalocyanine (\ce{H2Pc}) molecules with a central cavity are able to incorporate Ag atoms from an Ag(110) surface thus creating silver-phthalocyanine (AgPc).
The reaction was investigated by means of scanning tunneling microscopy (STM) under ultrahigh vacuum, and the metalation of \ce{H2Pc} at the interface was confirmed with X-ray photoelectron spectroscopy.
Three different kinds of molecules were found on the surface that are assigned to \ce{H2Pc}, the corresponding dehydrogenated molecules (Pc) and AgPc. The relative amounts of Pc and AgPc increase with increasing annealing temperature. We suggest that the reaction with Ag atoms from the steps of the surface occurs favorably only for already dehydrogenated molecules; thus, the metalation of \ce{H2Pc} is likely limited by the heat-induced dehydrogenation. Density functional theory simulations of the reaction path are presented to corroborate this hypothesis.

\end{abstract} 

\end{frontmatter}

\section{Introduction}

During the past decade, the on-surface metalation of ordered molecular layers of prototypical porphyrin\cite{Gottfried2006, Buchner2007, Kretschmann2007, Auwaerter2007, Buchner2008, Chen2010, Santo2012, Li2012} and phthalocyanine\cite{Bai2008, Song2010} molecules was developed and has gained considerable interest. This is achieved by evaporating metal atoms onto the surface after or before the deposition of molecules under ultrahigh-vacuum (UHV) conditions.
Various metals were tested, and the reaction kinetics studied, thus showing that depending on the kind of metal atom annealing at a high temperature might be necessary for incorporation into the molecule.\cite{Shubina2007}

To our knowledge, one of the first reports of the direct reaction with the atoms from the metal surface (``self-metalation'') appeared for porphyrin molecules on a Ag electrode; however, it was not perceived to be a new process for bottom-up synthesis of metal complexes directly by atoms from the metal substrate.\cite{Cotton1982}
Later, it was found that various porphyrin derivates readily reacted with available adatoms on a Cu surface and formed the corresponding metalated species.\cite{Gonzalez-Moreno2011, Veld2008, Diller2012} This reaction can already start at low temperatures where diffusing adatoms become available by detachment from the step edges of the surface.\cite{Gonzalez-Moreno2011}
Annealing at elevated temperatures drastically speeds up the reaction, and dependence of the reaction probability on the coverage of molecules\cite{Roeckert2014} and surface reconstruction\cite{Nowakowski2013} were also found recently. Self-metalation was found not only for Cu surfaces but also for other reactive surfaces such as Fe and Ni.\cite{Goldoni2012}

Various modifications of metal-free phthalocyanine and porphyrin molecules were often investigated on single crystals of noble metals like Au or Ag, most often on the close-packed (111) surface; however, the inclusion of these surface atoms in the molecular macrocycle has never been observed with scanning tunneling microscopy (STM) to our knowledge. Thus, it was believed that in this case a self-metalation on the surface is not possible because noble metal surfaces are quite inert.
Nevertheless, such molecules with coordinated silver or gold ions like silver phthalocyanine\cite{Gupta2007} or silver porphyrin\cite{Karweik1974, Karweik1976, Fukuzumi2008} are already long known and can be synthesized in solution.
Also, a forced metalation of a single phthalocyanine on Ag(111) supposedly with Ag atoms by pushing an atom at the tip of a STM into the previously dehydrogenated central cavity of the macrocycle was demonstrated recently.\cite{Sper2011}

Two questions arise: whether it is also possible to self-metalate molecules with Ag atoms directly from a crystalline Ag surface and why this was not observed in STM studies so far.
The specific surface orientation can be decisive for initiating reactions.\cite{Walch2010} Thus, a more reactive orientation of the surface plane of Ag could lead to success that was not possible on the close-packed Ag(111) surface.
Indeed, it was found using surface-enhanced resonance Raman spectroscopy that a free-base porphyrin macrocycle adsorbed on the roughened surface of Ag colloidal nanoparticles may incorporate a metal ion.\cite{Hanzlikova1998, Simakova2014} After deposition of 0.1 nm of Ag on top of a thick film of free-base phthalocyanine (\ce{H2Pc}), a shift of the HOMO and LUMO was reported.\cite{Gorgoi2006} This effect was explained by charge transfer to the deposited Ag atoms and the interaction of \ce{H2Pc} + Ag atoms described as chemisorption, whereas CuPc + Ag corresponded to physisorption.\cite{Gorgoi2006} We suggest that this strong interaction of \ce{H2Pc} could also have been due to a metalation reaction.

In summary, there are several reports for \textit{in situ} metalation reactions of porphyrins with Ag atoms and indications for this reaction with \ce{H2Pc}.
In this paper, we will show clear evidence by X-ray photoelectron spectroscopy (XPS) and STM that self-metalation is indeed also possible for \ce{H2Pc} on a Ag(110) surface. In the last part of this paper, density functional theory calculations (DFT) of the reaction path are presented and discussed.

\section{Experimental details}
\subsection{Sample preparation}

The Ag(110) single crystals were carefully cleaned under UHV according to the standard procedure by cycles of Ar$^+$-sputtering at an energy of 500\,eV and annealing at 500$\,^{\circ}$C for 1\,h. The quality and cleanliness of the prepared substrate was verified by STM or XPS, respectively, prior to the deposition of \ce{H2Pc} molecules (Sigma-Aldrich, purity of $>$ 97\,\%). The molecules were filled into a Knudsen cell in a separate preparation chamber and further purified with gradient sublimation by heating to a temperature slightly below the sublimation temperature ($\approx 350\,^{\circ}$C) under UHV. During the deposition of \ce{H2Pc}, the substrate was held at room temperature.
The thickness of the layer was estimated with a quartz microbalance for STM and by the attenuation of the substrate peaks (Ag 3d) for XPS.\cite{Seah1979} A monolayer (ML) of the molecular adsorbate is defined as the minimal thickness where the substrate surface is completely covered with a close-packed layer of molecules ($\approx$ 0.3\,nm).

For the STM experiments, the sample was annealed by resistive heating of the sample holder and the temperature was measured close to the sample plate on which the Ag(110) crystal was clamped.
For XPS, a different Ag(110) single crystal, which fits in the special sample holder used, was provided by the Material Science beamline. In this case, annealing of the sample was carried out by heating a wire that was wound directly around the Ag(110) crystal and held it in place; the thermocouple was fixed at its backside to achieve an accurate temperature measurement. In each annealing step, the temperature was increased slowly up to the designated maximum temperature at which point it was held for 10\,min and then cooled down to room temperature for the measurements.

\subsection{Methods}

The STM experiments were carried out with a variable temperature STM from Omicron. Measurements were performed at room-temperature if not noted otherwise. The base pressure in the UHV chamber with the STM was in the range of $10^{-10}$\,mbar. Electrochemically etched tungsten tips were used in the STM. All STM images were measured in constant current mode with a tunneling current of 100\,pA. Positive bias voltages correspond to tunneling to the sample, negative bias from the sample to the tip. STM images were processed with the WSxM software,\cite{Horcas2007} whereby slight filtering was applied to remove noise.

X-ray photoelectron spectroscopy of a thick layer ($\approx$ 7\,nm) of \ce{H2Pc} was measured with a Mg K$_\alpha$ source. The XPS experiments for the thin layers were carried out with synchrotron radiation at the Material Science beamline at Elettra (Trieste, Italy). A Phoibos photoelectron spectrometer was used. The photoelectrons were collected in the surface normal
direction with a photon incidence angle of 30$\,^{\circ}$. The binding energies were corrected relative to the Ag 3d$\_{5/2}$ substrate peak (368.2\,eV) or for the thick layer, for which Ag 3d was not visible, to the Au 4f$\_{7/2}$ (84.0\,eV) peak of a clean reference gold foil.
Core-level spectra were fitted with a Shirley and a linear background, and peak functions of Voigt shape using an identical width for all peaks. 

DFT calculations were carried out using the grid-based projector augmented wave method (GPAW).\cite{Enkovaara2010} The RPBE exchange-correlation functional with a pairwise correction for the dispersion interaction [vdW(TS)]\cite{Tkatchenko2009}, and a real-space grid with spacing of $h = 0.2$\,\AA{} were used. The simplified model consists of a (110)-oriented substrate slab of three layers of Ag, with the two lowest layers fixed during optimization, an Ag adatom, and the \ce{H2Pc} (or Pc) molecule that is centered on a bridge position on top of an atomic Ag row. The molecule is chosen to be rotated by 23$\,^{\circ}$ relative to the direction of the row ($[1\bar{1}0]$) as found by STM. The simulation cell has periodic boundary conditions in $x$ and $y$ and zero boundary in $z$ direction with a vacuum of in total 1.3\,nm.
The reaction path of the metalation was modeled by a ``nudged elastic band'' (NEB), which was consecutively relaxed first in the LCAO mode with the standard double-$\zeta$-polarized (dzp) basis set of atomic orbitals for the representation of the wave functions and at the end switched to the finite difference mode (grid-only) for fine-tuning.\cite{Larsen2009}

\section{Results and Discussion}
\subsection{XPS investigations}

\begin{figure}[tb]
\centering
\includegraphics[width=0.47\textwidth]{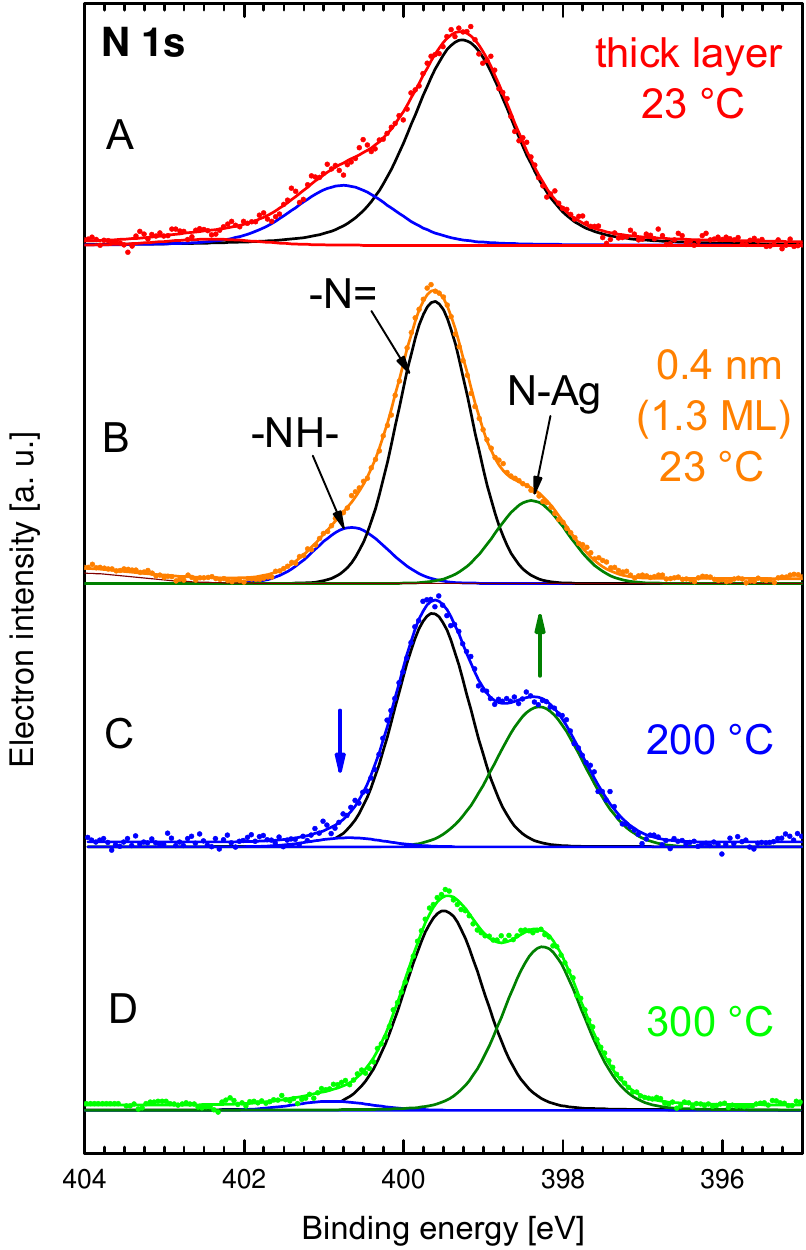}
\caption{Evolution of the N 1s core level spectra of \ce{H2Pc} on Ag(110). (A) thick layer ($\approx$7\,nm), $E\_{exc} = 1253.6\,$eV (Mg K$_\alpha$) B-D: Initially $\approx$ 0.4\,nm (1.3\,ML), $E\_{exc} = 485\,$eV. The maximum temperature during annealing is shown at each curve. Arrows indicate an increase/decrease of the corresponding peak component.}
\label{fig:PES_N1s}
\end{figure}

Changes in the 1s core level of the nitrogen atoms are a good indication for reactions occurring at the center of the molecule (Fig.~\ref{fig:PES_N1s}). The N 1s signal of \ce{H2Pc} molecules consists of two components for pyrrolic (-NH-) and iminic (-N=) nitrogen with a peak area ratio of 1 : 3, as can be seen for a thick layer in Fig.~\ref{fig:PES_N1s}A. At a coverage of slightly more than one monolayer (Fig.~\ref{fig:PES_N1s}B) an additional peak at lower binding energy of circa 398.3\,eV is observed. This species of N atoms can be attributed to metalated phthalocyanine, wherein the four central N atoms coordinate to a metal ion. The metal atom inside the molecule is usually positively charged (e.g. oxidation state 2+) and donates electron density to the surrounding N atoms thereby leading to a shift to lower N 1s binding energy. Because no metal atoms were deposited before or after the adsorption of the molecules and the pristine substrate was confirmed to be clean and free of impurities, the aforementioned metal can only be Ag originating from the Ag(110) substrate.
A N 1s binding energy of 397.6\,eV for Ag(II) octaethylporphyrin\cite{Karweik1974} and 398.3\,eV\cite{Karweik1974} or 398.5\,eV\cite{Karweik1976} for Ag(II)-tetraphenylporphyin were reported before, which are similar to the energy of the peak measured here that is attributed to N-Ag bonds.

Notably, XPS indicates that around one-third of the molecules were already metalated at room temperature, which indicates a low reaction barrier for the metalation. On the anisotropic Ag(110), Ag atoms already begin to detach from step edges at temperatures over 175\,K and subsequently diffuse over the surface, preferentially along the Ag rows ($[1\bar{1}0]$ direction).\cite{Morgenstern1999} The availability of Ag atoms necessary for the reaction is significantly higher on Ag(110) than on the close-packed Ag(111) surface at room temperature. This might be the reason why on Ag(111) this kind of metalation with surface atoms has never been observed.

Annealing (Figures~\ref{fig:PES_N1s}C and D) led to a large decrease of the peak at 400.7\,eV (-NH-) and an increase of the peak at 398.3\,eV, which means that more \ce{H2Pc} molecules were metalated at increased temperature. For AgPc, the amount of iminic (-N=) and N-Ag nitrogen atoms is equal. Therefore, it can be seen that after the sample was heated to 300$\,^{\circ}$C, around 82\% of \ce{H2Pc} molecules seemingly reacted with surface atoms.
No significant change with annealing could be found in the Ag 3d core level, likely because the contribution from the substrate clearly dominates the signal.

\subsection{STM results of the molecular adlayer}

\begin{figure}[!tb]
\centering
\includegraphics[width=0.47\textwidth]{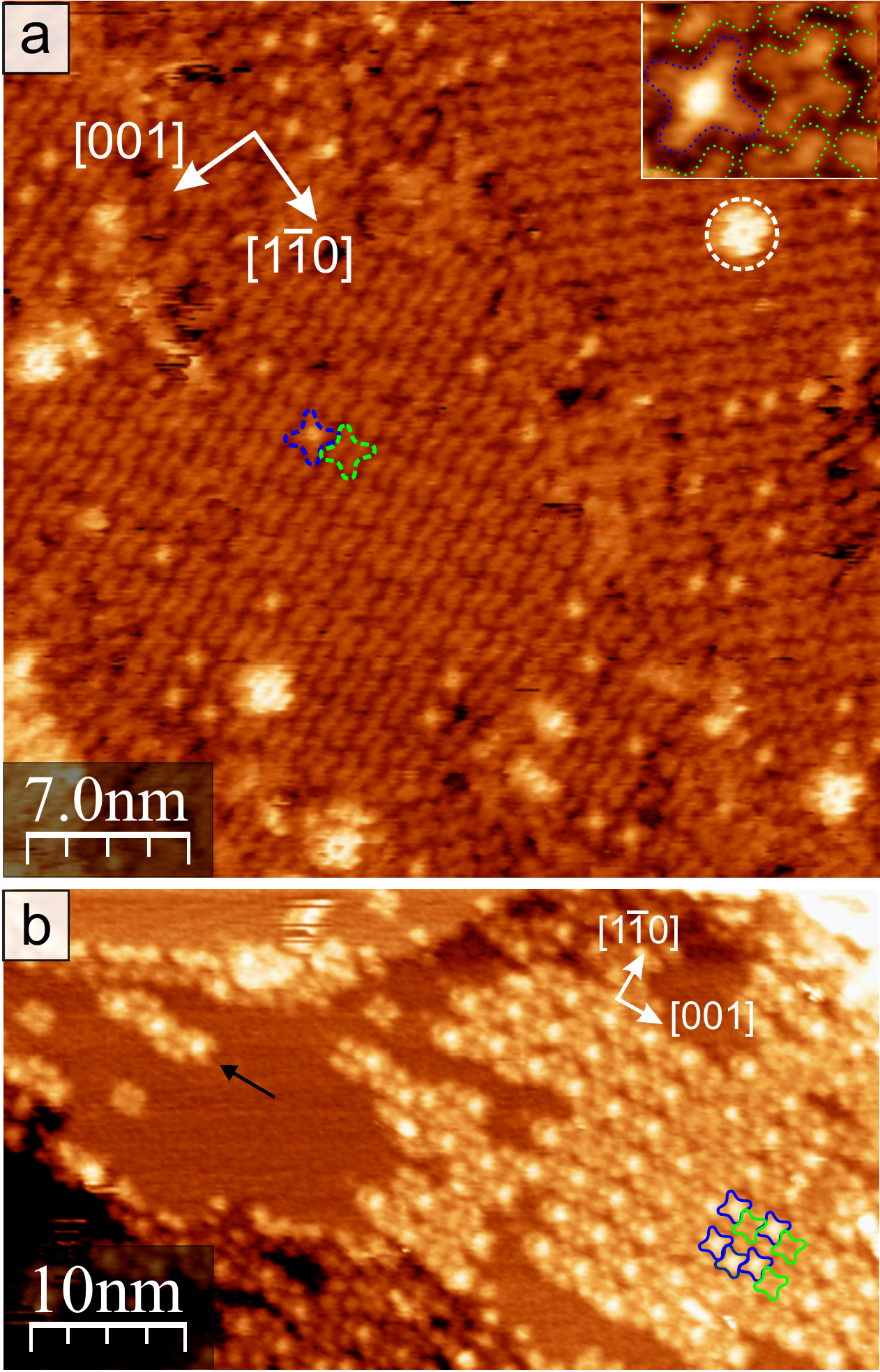}
\caption{(a) STM image of monolayer coverage of \ce{H2Pc} on Ag(110) annealed to $\approx$ 100$\,^{\circ}$C (36.8\,nm $\times$ 36.8\,nm, $U = -0.8$\,V, $I = 100$\,pA, measured at room temperature). \ce{H2Pc} and a metalated phthalocyanine are indicated exemplarily with green and blue, respectively, and shown magnified in the inset. A single \ce{H2Pc} molecule adsorbed on top of the molecular monolayer is marked with a white circle. The directions of the lattice vectors of the underlying Ag(110) substrate are shown by the white arrows.
(b) STM of submonolayer coverage annealed to $\approx$ 200$\,^{\circ}$C (55\,nm $\times$ 30\,nm, $U = -1.5$\,V, $I = 100$\,pA, at 38\,K). An alternating arrangement of \ce{H2Pc} and AgPc molecules is formed (arrow). Step edges of Ag(110) are partially decorated with molecules.}
\label{fig:STM_metalate}
\end{figure}

Microscopic evidence of metalation is shown in Fig.~\ref{fig:STM_metalate}. \ce{H2Pc} molecules in direct interaction with the Ag(110) substrate appear in a crosslike shape with a depression in the center (framed green). Single \ce{H2Pc} molecules adsorbed on top of the monolayer (Fig.~\ref{fig:STM_metalate}(a), white circle) are decoupled from the substrate and have a more detailed appearance with two protrusions on each lobe of the molecule. Additionally, crosslike molecules with a central protrusion, which are assigned to AgPc molecules, were observed (Fig.~\ref{fig:STM_metalate}, framed in blue). This central protrusion (apparent height of $\approx 0.1\,$nm) is seemingly mostly an electronic effect due to d orbitals with extension in the $z$ direction (d$\_{xz}$, d$\_{yz}$) because it is not clearly visible at all bias voltages.

Annealing of the sample again resulted in an increasing amount of AgPc relative to \ce{H2Pc} molecules. At all annealing steps and especially at room temperature, far less metalated molecules were found in STM than the XPS results suggested. The reason for this could be a different roughness of the Ag(110) surfaces of the different crystals used for the respective experiments, i.e., the size of flat terraces and number of steps on the surface. More steps lead to a higher availability of diffusing Ag atoms and, subsequently, likely a higher probability for metalation of the molecules.
It should also be noted that STM is a local technique and mostly flat areas of the surface were chosen for imaging instead of areas with many steps. In contrast, XPS averages over nearly the whole sample, which likely leads to the measured higher ratio of AgPc to \ce{H2Pc} compared to that measured by STM.

The amount of metalated molecules after annealing seemed to be higher at submonolayer coverage (Fig.~\ref{fig:STM_metalate}(b)) compared to that at a full monolayer, which might be due to easier diffusion of Ag adatoms to the molecules. It was also reported previously that the rate of metalation can strongly depend on the coverage of deposited molecules.\cite{Roeckert2014} A different coverage can lead to different adsorbate structures, which might hinder or favor the diffusion of adatoms. Besides the scenario of metalation with a diffusing adatom, there is also the possibility that a metal atom is directly taken out of the surface underneath the center of the molecule. This, however, creates a vacancy at the surface which is thermodynamically very unfavorable.\cite{Goldoni2012}

Interestingly, after annealing a coverage of around half a monolayer of \ce{H2Pc} on Ag(110), the molecules sometimes rearranged into a new two-dimensional structure consisting of molecular rows along [001] of alternating and zigzag displaced \ce{H2Pc} and AgPc molecules (Fig.~\ref{fig:STM_metalate}(b)). This binary structure is likely due to different favorable adsorption positions of \ce{H2Pc} and AgPc on the Ag rows of Ag(110) or to an electrostatic repulsive interaction if charge in the substrate under the AgPc molecules is pushed away or accumulated by the central metal ion.

\begin{figure*}[!tb]
\centering
\includegraphics[width=0.68\textwidth]{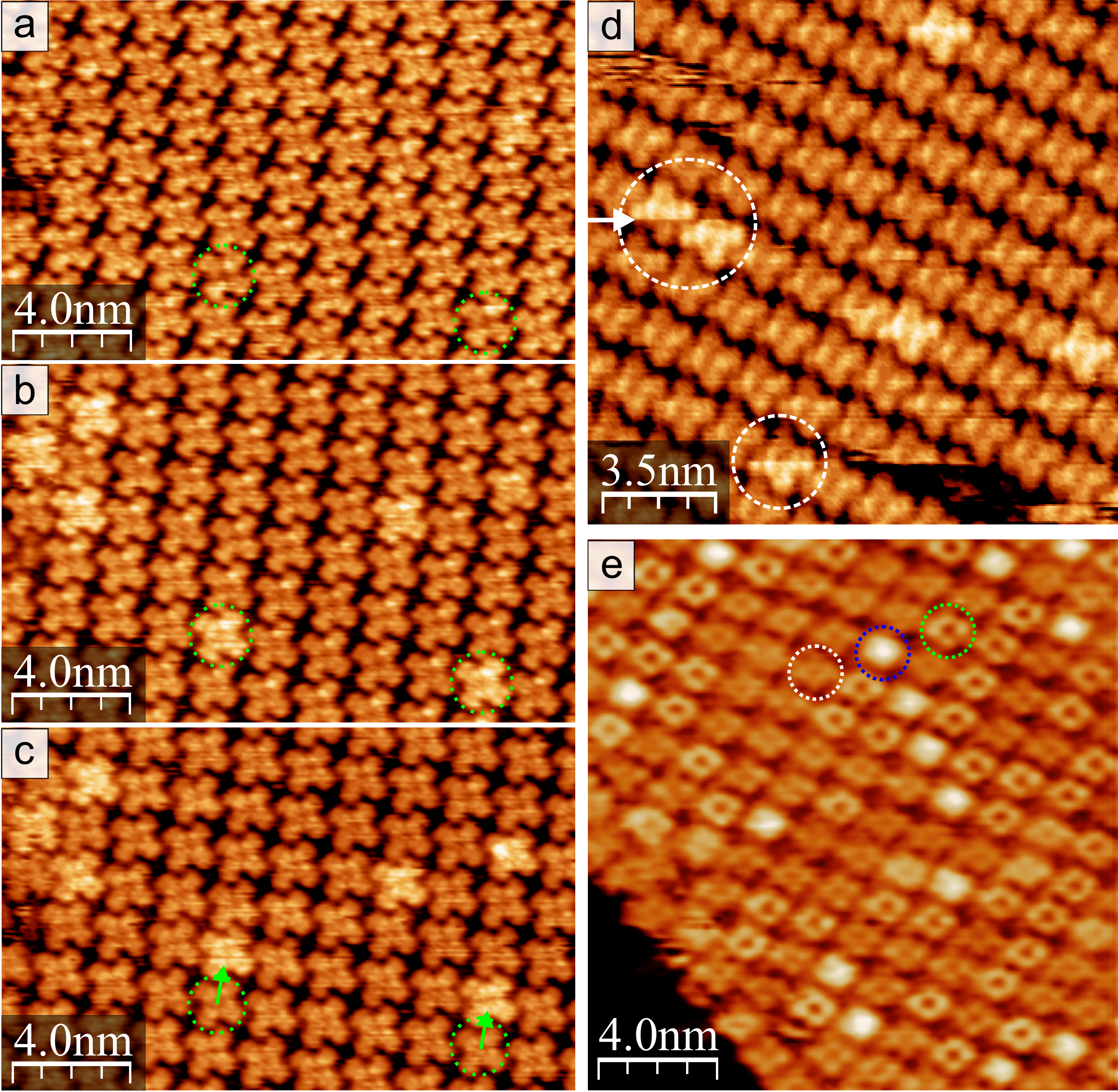}
\caption{(a)-(c) Series of STM images of the same scan area showing the two species of metal-free phthalocyanine. ((a) $U = +1.2$\,V, (b) and (c) $U = +0.2$\,V, $I = 1$\,nA) The brighter contrast of two molecules visible at low bias voltage seemingly moved inside the close-packed layer to adjacent positions (marked by green arrows). 
(d) Contrast change of individual molecules (white circles) during scanning as a result of proposed tip-induced dehydrogenation of \ce{H2Pc}, and hydrogen transfer and hydrogenation of neighboring Pc molecules. The slow scanning direction was from bottom to top, and fast scanning was from left to right. ($U = +0.2$\,V, $I = 100$\,pA)
(e) STM image of the three different molecular species with slightly different resolution and contrast: dim molecules \ce{H2Pc} (white circle); molecules with larger central depression (Pc, green circle); molecules with central protrusion (AgPc, blue circle) ($U = -0.8$\,V, $I = 100$\,pA)}
\label{fig:deprot}
\end{figure*}

Although \ce{H2Pc} shows a small central depression and the metalated Pc shows a protrusion in STM, after annealing a large amount of a third, flat molecular appearance was observed in the monolayer (Fig.~\ref{fig:deprot}). The brighter appearance (or slightly different apparent height) can only be observed at small positive or negative bias voltage, whereas at higher bias this species usually cannot be discriminated clearly from \ce{H2Pc} (Fig.~\ref{fig:deprot}(a)). In the image of Fig.~\ref{fig:deprot}(e) (sample annealed to $\approx$ 100$\,^{\circ}$C), the STM contrast for this tip and at $U = -0.8$\,V again shows the \ce{H2Pc} molecules with a shallow central depression (white circle) and AgPc with a big central protrusion (blue circle; however, the second species is imaged now with a large central depression (green circle).

One explanation for this species is an intermediate state in the process of metalation.\cite{Kretschmann2007, Li2012}. If Ag atoms diffuse (preferably along silver rows) underneath the molecular layer, then one could be trapped under the center of the metal-free phthalocyanine without being lifted up inside the molecule. For this step and the dissociating of the two central hydrogen atoms an energy barrier should be overcome, which will be discussed in more detail later. Ag atoms underneath the molecular plane and partially bonded to the N atoms should also lead to a shift in the N 1s core-level spectra and, therefore, to a contribution to the feature contributed to N-Ag.
Against this explanation is the fact that the bright molecules seldom seem to move in the layer, as can be seen when Fig.~\ref{fig:deprot}(b) and (c) are compared or even change the contrast between dark and bright during the scan by the influence of the tip (Fig.~\ref{fig:deprot}(d)).

Notably, a modification of \ce{H2Pc} (17\%) on Ag(111) was reported before by Bai and co-workers\cite{Bai2008} The origin of the second kind of molecules was not clear; however, both species could be metalated by deposition of Fe atoms. Another more likely explanation would thus be that these molecules correspond to dehydrogenated phthalocyanine (Pc) which has a slightly different local density of states than \ce{H2Pc} similar to what was previously found for a metal-free tetra(\textit{p}-hydroxyphenyl)porphyrin.\cite{Smykalla2014a} The higher contrast of dehydrogenated phthalocyanine at voltages around the Fermi energy might also indicate hybridization with the surface state of Ag(110).
The in-layer movement and change of contrast shown in Fig.~\ref{fig:deprot} can thus be understood by transfer of \ce{H2} between \ce{H2Pc} and Pc molecules induced by the influence of the STM tip. For example, in the upper white circle in Fig.~\ref{fig:deprot}(d), during the scan line (slow scan direction from bottom to top, fast scan direction from left to right) the left molecule is dehydrogenated and turns bright, and two hydrogen atoms are transferred by the tip to the adjacent bright Pc molecule which is subsequently hydrogenated again to the dimmer \ce{H2Pc}.
The occurrence and increasing amount of dehydrogenated phthalocyanine is due to annealing of the sample\cite{Roeckert2014} and the catalytic activity of Ag(110).

For dehydrogenated phthalocyanine molecules, a metalation with Ag atoms is favorable to saturate again the pyrrolic N atoms in the center. A two-step metalation process with a higher reaction barrier is in this case not required. We propose, therefore, that \ce{H2Pc} does not react directly with Ag but rather that a preceding hydrogen dissociation is required.\cite{Sper2011}
Also, metalated molecules are often not preferably found near step edges of Ag(110) but at random positions inside the molecular domains, which supports the assumption that mostly the dehydrogenated molecules capture and incorporate diffusing Ag atoms. The relative small amount of AgPc molecules observed by STM even after extensive annealing indicates that the metalation reaction is not very efficient and the close-packed arrangement of the molecules likely hinders the diffusion of adatoms to the dehydrogenated phthalocyanine molecules.
Also, self-metalation with Ag was not found for metal-free tetra(\textit{p}-hydroxyphenyl)porphyrin on Ag(110).\cite{Smykalla2014b} The reason for this is probably the slightly larger distance of the molecular macrocycle from the Ag surface because of the elevation by the tilted phenyl rings or the partial hydrogenation of the central nitrogen atoms after annealing to high temperatures, which prevents a metalation.\cite{Smykalla2015a}

\subsection{DFT simulations of the reaction path}

Density functional theory optimizations of the AgPc molecule indicate that it has the typical planar D$\_{4h}$ symmetry. The Ag ion fits inside the central cavity of the macrocycle, and no shuttlecock-like deformation such as that in the case of SnPc\cite{Toader2011a} occurs in gas phase or after adsorption.
We also obtained this flat geometry with different basis sets and xc-functionals (vdW-DF2, PBE, B3LYP) in gas phase, and only a small deformation was found for the relaxed molecule on the Ag slab.
Also, it was not possible in STM to ``switch'' the metalated molecules by manipulation with the tip into a geometry with metal ion pointing toward the surface, which seems to be different from the report by Sperl \textit{et al.}\cite{Sper2011} but corroborates a flat geometry of the AgPc molecule.

\begin{figure*}[!tb]
\centering
\includegraphics[width=0.85\textwidth]{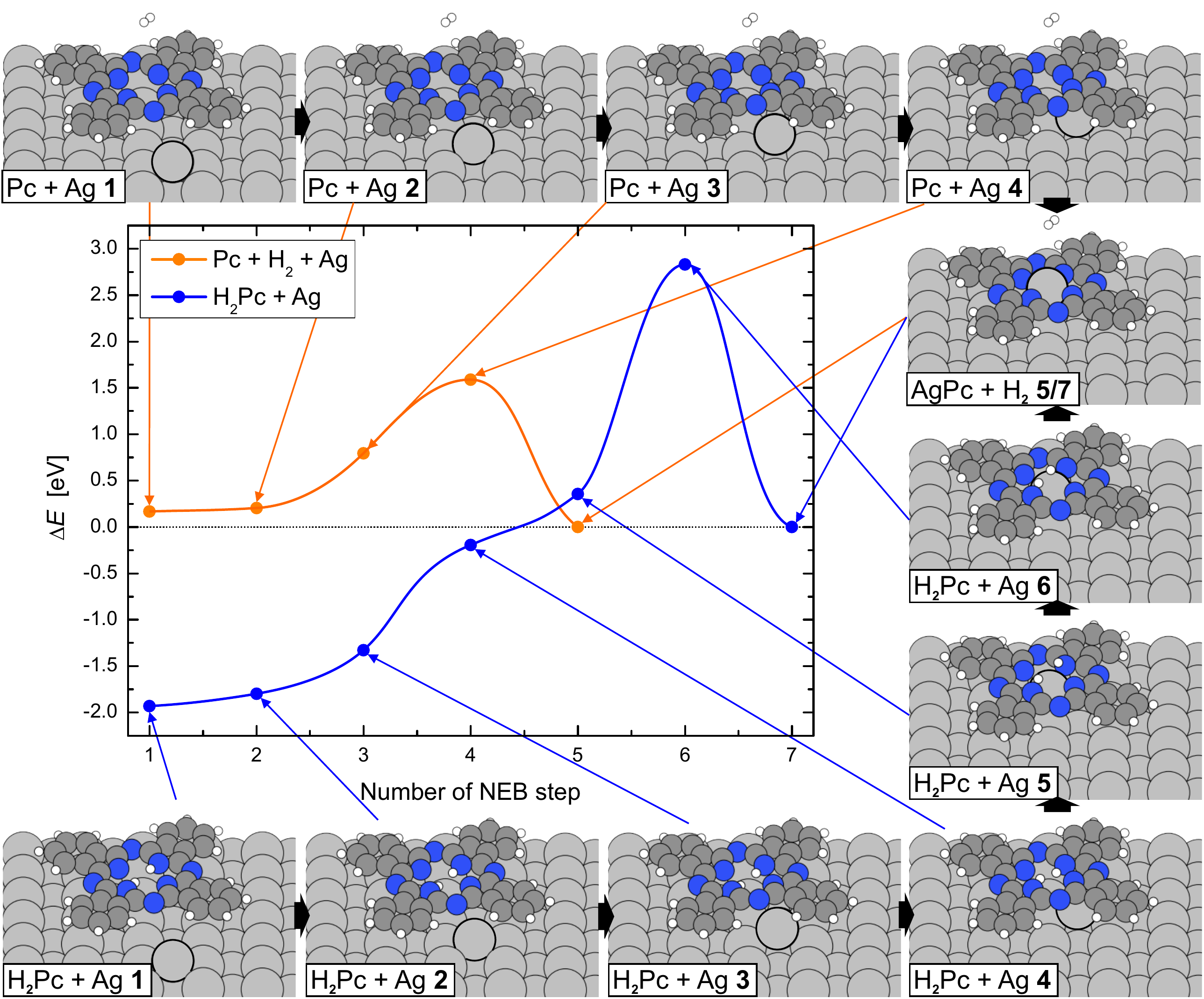}
\caption{Nudged elastic band paths for the metalation reaction of \ce{H2Pc} and the dehydrogenated form Pc with a diffusing Ag adatom (slightly highlighted) on Ag(110). The change of the atomic positions at each step is shown, and point by arrows to the corresponding change of the total energy relative to the final step of AgPc + H$\_2$.
The elements are colored as follows: silver, bright gray; carbon, dark gray; nitrogen, blue; and hydrogen atoms, white.}
\label{fig:AgPc_neb}
\end{figure*}

\newpage
To our knowledge, only gas phase calculations for the reaction path of a metalation were so far carried out because the adsorption on a surface leads to many possible paths for diffusion and reaction of the metal ion with the molecule. Furthermore, it is not clear if the metal ion can move underneath the molecule (and thereby lift it) to the central cavity.
The gas phase calculations for the metalation of porphine molecules with different metal atoms showed that the metal ion is coordinated by the four nitrogen atoms with the hydrogen atoms still bonded to two N atoms.\cite{Shubina2007} Thereafter, the two aminic H atoms are successively transferred to the metal ion where they combine and finally detach as molecular hydrogen.

The calculated reaction paths of the metalation of \ce{H2Pc} (blue trace) or Pc (orange trace) with Ag are shown in Fig.~\ref{fig:AgPc_neb}. First, the Ag adatom moves along the trench from the Ag substrate rows, which is the path with the lowest diffusion barrier on Ag(110). It then has to move underneath the molecule (steps 1 to 4) to reach the center of the molecule (step 5). Thereby, this part of the molecule is locally lifted, which requires additional energy. The metalation of an already dehydrogenated \ce{H2Pc} (Pc, orange trace, Fig.~\ref{fig:AgPc_neb}) requires only this first energy barrier for the diffusion of Ag to the molecular cavity. On the other hand, in step 5 of \ce{H2Pc} + Ag, the two hydrogen bonded at the central N atoms hinder the immediate metalation. Thus, the Ag atom is coordinated by the N atoms but is still below the molecular plane and the two H atoms are bent away from the surface. Another energy barrier has to be overcome to detach one H from the respective N atom, which can then move over the Ag atom to the second hydrogen (step 6). Both H atoms can then form H$_2$ and desorb, whereas the Ag atom is lifted up in the plane of the molecule (step 7).

The height of the total energy barrier is 109\,kcal/mol (4.7\,eV) for \ce{H2Pc} + Ag and 33 kcal/mol (1.4\,eV) for Pc + Ag. The metalation of \ce{H2Pc} with Ag is highly endothermic, which could be the reason why it was usually not observed previously.
For comparison, the metalation of the similar porphine molecule with Fe, Co, Cu, and Zn was found to be always exothermic, with the enthalpy and barrier height depending on the specific metal.\cite{Shubina2007} For the metalation of tetraphenylporphyrin on Ni(111) with Ni surface atoms, a gain in energy of 0.89\,eV was calculated, and in the reported XPS experiment, the reaction already readily proceeded at room temperature on Ni(111).\cite{Goldoni2012}

\section{Conclusions}

We showed that the reaction of metal-free phthalocyanine molecules with Ag atoms from a Ag(110) surface upon adsorption already occurs at a small percentage already at room temperature, which leads to the metalation of the molecules. Our DFT simulations indicate that the metalation of \ce{H2Pc} is endothermic and has a high energy barrier. However, for the dehydrogenated species Pc, the reaction has a significantly lower barrier and is thermodynamically favorable. The metalation of \ce{H2Pc} to AgPc on Ag(110) thus likely consists of two separate reactions that are the dissociation of the central hydrogen atoms by annealing and the diffusion of Ag atoms from step edges to dehydrogenated molecules where they are coordinated by the four N atoms.

Our finding shows that the orientation of the substrate surface can be crucial for self-metalation reactions because it was not observed for \ce{H2Pc} on Ag(111). Furthermore, for electronic devices applying ultrathin films of \ce{H2Pc}, Ag contacts could also lead to reactions to AgPc and consequently alter the desired properties.

\section*{Acknowledgements}

This work has been financially supported by the Deutsche Forschungsgemeinschaft (DFG) through the Research Unit FOR 1154. Photoelectron spectroscopy was carried out at the Material Science beamline at the synchrotron Elettra (Trieste, Italy), where we thank Martin Vondr$\acute{\mathrm{a}}$\v{c}ek for technical assistance. Computational resources were provided by the ``Chemnitzer Hochleistungs-Linux-Cluster'' (CHiC) at the Technische Universit\"at Chemnitz.

\bibliography{AgPc}

\providecommand*\mcitethebibliography{\thebibliography}
\csname @ifundefined\endcsname{endmcitethebibliography}
  {\let\endmcitethebibliography\endthebibliography}{}
\begin{mcitethebibliography}{38}
\providecommand*\natexlab[1]{#1}
\providecommand*\mciteSetBstSublistMode[1]{}
\providecommand*\mciteSetBstMaxWidthForm[2]{}
\providecommand*\mciteBstWouldAddEndPuncttrue
  {\def\EndOfBibitem{\unskip.}}
\providecommand*\mciteBstWouldAddEndPunctfalse
  {\let\EndOfBibitem\relax}
\providecommand*\mciteSetBstMidEndSepPunct[3]{}
\providecommand*\mciteSetBstSublistLabelBeginEnd[3]{}
\providecommand*\EndOfBibitem{}
\mciteSetBstSublistMode{f}
\mciteSetBstMaxWidthForm{subitem}{(\alph{mcitesubitemcount})}
\mciteSetBstSublistLabelBeginEnd
  {\mcitemaxwidthsubitemform\space}
  {\relax}
  {\relax}

\bibitem[Gottfried et~al.(2006)Gottfried, Flechtner, Kretschmann, Lukasczyk,
  and Steinr\"{u}ck]{Gottfried2006}
Gottfried,~J.~M.; Flechtner,~K.; Kretschmann,~A.; Lukasczyk,~T.;
  Steinr\"{u}ck,~H.-P. {Direct Synthesis of a Metalloporphyrin Complex on a
  Surface}. \emph{J. Am. Chem. Soc.} \textbf{2006}, \emph{128},
  5644--5645\relax
\mciteBstWouldAddEndPuncttrue
\mciteSetBstMidEndSepPunct{\mcitedefaultmidpunct}
{\mcitedefaultendpunct}{\mcitedefaultseppunct}\relax
\EndOfBibitem
\bibitem[Buchner et~al.(2007)Buchner, Schwald, Comanici, Steinr\"{u}ck, and
  Marbach]{Buchner2007}
Buchner,~F.; Schwald,~V.; Comanici,~K.; Steinr\"{u}ck,~H.-P.; Marbach,~H.
  {Microscopic Evidence of the Metalation of a Free-Base Porphyrin Monolayer
  with Iron}. \emph{ChemPhysChem} \textbf{2007}, \emph{8}, 241--243\relax
\mciteBstWouldAddEndPuncttrue
\mciteSetBstMidEndSepPunct{\mcitedefaultmidpunct}
{\mcitedefaultendpunct}{\mcitedefaultseppunct}\relax
\EndOfBibitem
\bibitem[Kretschmann et~al.(2007)Kretschmann, Walz, Flechtner, Steinr\"{u}ck,
  and Gottfried]{Kretschmann2007}
Kretschmann,~A.; Walz,~M.-M.; Flechtner,~K.; Steinr\"{u}ck,~H.-P.;
  Gottfried,~J.~M. {Tetraphenylporphyrin Picks up Zinc Atoms from a Silver
  Surface}. \emph{Chem. Commun.} \textbf{2007}, 568--570\relax
\mciteBstWouldAddEndPuncttrue
\mciteSetBstMidEndSepPunct{\mcitedefaultmidpunct}
{\mcitedefaultendpunct}{\mcitedefaultseppunct}\relax
\EndOfBibitem
\bibitem[Auw\"{a}rter et~al.(2007)Auw\"{a}rter, Weber-Bargioni, Brink, Riemann,
  Schiffrin, Ruben, and Barth]{Auwaerter2007}
Auw\"{a}rter,~W.; Weber-Bargioni,~A.; Brink,~S.; Riemann,~A.; Schiffrin,~A.;
  Ruben,~M.; Barth,~J.~V. {Controlled Metalation of Self-Assembled Porphyrin
  Nanoarrays in Two Dimensions}. \emph{ChemPhysChem} \textbf{2007}, \emph{8},
  250--254\relax
\mciteBstWouldAddEndPuncttrue
\mciteSetBstMidEndSepPunct{\mcitedefaultmidpunct}
{\mcitedefaultendpunct}{\mcitedefaultseppunct}\relax
\EndOfBibitem
\bibitem[Buchner et~al.(2008)Buchner, Flechtner, Bai, Zillner, Kellner,
  Steinr\"{u}ck, Marbach, and Gottfried]{Buchner2008}
Buchner,~F.; Flechtner,~K.; Bai,~Y.; Zillner,~E.; Kellner,~I.;
  Steinr\"{u}ck,~H.-P.; Marbach,~H.; Gottfried,~J.~M. {Coordination of Iron
  Atoms by Tetraphenylporphyrin Monolayers and Multilayers on Ag(111) and
  Formation of Iron-Tetraphenylporphyrin}. \emph{J. Phys. Chem. C}
  \textbf{2008}, \emph{112}, 15458--15465\relax
\mciteBstWouldAddEndPuncttrue
\mciteSetBstMidEndSepPunct{\mcitedefaultmidpunct}
{\mcitedefaultendpunct}{\mcitedefaultseppunct}\relax
\EndOfBibitem
\bibitem[Chen et~al.(2010)Chen, Feng, Zhang, Ju, Xu, Zhu, Gottfried, Ibrahim,
  Qian, and Wang]{Chen2010}
Chen,~M.; Feng,~X.; Zhang,~L.; Ju,~H.; Xu,~Q.; Zhu,~J.; Gottfried,~J.~M.;
  Ibrahim,~K.; Qian,~H.; Wang,~J. {Direct Synthesis of Nickel(II)
  Tetraphenylporphyrin and Its Interaction with a Au(111) Surface}. \emph{J.
  Phys. Chem. C} \textbf{2010}, \emph{114}, 9908--9916\relax
\mciteBstWouldAddEndPuncttrue
\mciteSetBstMidEndSepPunct{\mcitedefaultmidpunct}
{\mcitedefaultendpunct}{\mcitedefaultseppunct}\relax
\EndOfBibitem
\bibitem[{Di Santo} et~al.(2012){Di Santo}, Sfiligoj, Castellarin-Cudia,
  Verdini, Cossaro, Morgante, Floreano, and Goldoni]{Santo2012}
{Di Santo},~G.; Sfiligoj,~C.; Castellarin-Cudia,~C.; Verdini,~A.; Cossaro,~A.;
  Morgante,~A.; Floreano,~L.; Goldoni,~A. {Changes of the Molecule-Substrate
  Interaction upon Metal Inclusion into a Porphyrin}. \emph{Chem. Eur. J.}
  \textbf{2012}, \emph{18}, 12619--12623\relax
\mciteBstWouldAddEndPuncttrue
\mciteSetBstMidEndSepPunct{\mcitedefaultmidpunct}
{\mcitedefaultendpunct}{\mcitedefaultseppunct}\relax
\EndOfBibitem
\bibitem[Li et~al.(2012)Li, Xiao, Shubina, Chen, Shi, Schmid, Steinr\"{u}ck,
  Gottfried, and Lin]{Li2012}
Li,~Y.; Xiao,~J.; Shubina,~T.~E.; Chen,~M.; Shi,~Z.; Schmid,~M.;
  Steinr\"{u}ck,~H.-P.; Gottfried,~J.~M.; Lin,~N. {Coordination and Metalation
  Bifunctionality of Cu with 5,10,15,20-Tetra(4-pyridyl)porphyrin: Toward a
  Mixed-Valence Two-Dimensional Coordination Network}. \emph{J. Am. Chem. Soc.}
  \textbf{2012}, \emph{134}, 6401--6408\relax
\mciteBstWouldAddEndPuncttrue
\mciteSetBstMidEndSepPunct{\mcitedefaultmidpunct}
{\mcitedefaultendpunct}{\mcitedefaultseppunct}\relax
\EndOfBibitem
\bibitem[Bai et~al.(2008)Bai, Buchner, Wendahl, Kellner, Bayer, Steinr\"{u}ck,
  Marbach, and Gottfried]{Bai2008}
Bai,~Y.; Buchner,~F.; Wendahl,~M.~T.; Kellner,~I.; Bayer,~A.;
  Steinr\"{u}ck,~H.-P.; Marbach,~H.; Gottfried,~J.~M. {Direct Metalation of a
  Phthalocyanine Monolayer on Ag(111) with Coadsorbed Iron Atoms}. \emph{J.
  Phys. Chem. C} \textbf{2008}, \emph{112}, 6087--6092\relax
\mciteBstWouldAddEndPuncttrue
\mciteSetBstMidEndSepPunct{\mcitedefaultmidpunct}
{\mcitedefaultendpunct}{\mcitedefaultseppunct}\relax
\EndOfBibitem
\bibitem[Song et~al.(2010)Song, Wang, Ning, Jia, Chen, Sun, Zhang, Xue, and
  Ma]{Song2010}
Song,~C.-L.; Wang,~Y.-L.; Ning,~Y.-X.; Jia,~J.-F.; Chen,~X.; Sun,~B.;
  Zhang,~P.; Xue,~Q.-K.; Ma,~X. {Tailoring Phthalocyanine Metalation Reaction
  by Quantum Size Effect}. \emph{J. Am. Chem. Soc.} \textbf{2010}, \emph{132},
  1456--1457\relax
\mciteBstWouldAddEndPuncttrue
\mciteSetBstMidEndSepPunct{\mcitedefaultmidpunct}
{\mcitedefaultendpunct}{\mcitedefaultseppunct}\relax
\EndOfBibitem
\bibitem[Shubina et~al.(2007)Shubina, Marbach, Flechtner, Kretschmann, Jux,
  Buchner, Steinr{\"u}ck, Clark, and Gottfried]{Shubina2007}
Shubina,~T.; Marbach,~H.; Flechtner,~K.; Kretschmann,~A.; Jux,~N.; Buchner,~F.;
  Steinr{\"u}ck,~H.-P.; Clark,~T.; Gottfried,~J. {Principle and Mechanism of
  Direct Porphyrin Metalation}. \emph{J. Am. Chem. Soc.} \textbf{2007},
  \emph{129}, 9476--9483\relax
\mciteBstWouldAddEndPuncttrue
\mciteSetBstMidEndSepPunct{\mcitedefaultmidpunct}
{\mcitedefaultendpunct}{\mcitedefaultseppunct}\relax
\EndOfBibitem
\bibitem[Cotton et~al.(1982)Cotton, Schultz, and {Van Duyne}]{Cotton1982}
Cotton,~T.~M.; Schultz,~S.~G.; {Van Duyne},~R.~P. {Surface-Enhanced Resonance
  Raman Scattering from Water-Soluble Porphyrins Adsorbed on a Silver
  Electrode}. \emph{J. Am. Chem. Soc.} \textbf{1982}, \emph{104},
  6528--6532\relax
\mciteBstWouldAddEndPuncttrue
\mciteSetBstMidEndSepPunct{\mcitedefaultmidpunct}
{\mcitedefaultendpunct}{\mcitedefaultseppunct}\relax
\EndOfBibitem
\bibitem[Gonz\'{a}lez-Moreno et~al.(2011)Gonz\'{a}lez-Moreno,
  S\'{a}nchez-S\'{a}nchez, Trelka, Otero, Cossaro, Verdini, Floreano,
  Ruiz-Bermejo, Garc\'{\i}a-Lekue, Mart\'{\i}n-Gago, and
  Rogero]{Gonzalez-Moreno2011}
Gonz\'{a}lez-Moreno,~R. et~al.  {Following the Metalation Process of
  Protoporphyrin IX with Metal Substrate Atoms at Room Temperature}. \emph{J.
  Phys. Chem. C} \textbf{2011}, \emph{115}, 6849--6854\relax
\mciteBstWouldAddEndPuncttrue
\mciteSetBstMidEndSepPunct{\mcitedefaultmidpunct}
{\mcitedefaultendpunct}{\mcitedefaultseppunct}\relax
\EndOfBibitem
\bibitem[{In't Veld} et~al.(2008){In't Veld}, Iavicoli, Haq, Amabilino, and
  Raval]{Veld2008}
{In't Veld},~M.; Iavicoli,~P.; Haq,~S.; Amabilino,~D.~B.; Raval,~R. {Unique
  Intermolecular Reaction of Simple Porphyrins at a Metal Surface Gives
  Covalent Nanostructures}. \emph{Chem. Commun.} \textbf{2008},
  1536--1538\relax
\mciteBstWouldAddEndPuncttrue
\mciteSetBstMidEndSepPunct{\mcitedefaultmidpunct}
{\mcitedefaultendpunct}{\mcitedefaultseppunct}\relax
\EndOfBibitem
\bibitem[Diller et~al.(2012)Diller, Klappenberger, Marschall, Hermann, Nefedov,
  W\"{o}ll, and Barth]{Diller2012}
Diller,~K.; Klappenberger,~F.; Marschall,~M.; Hermann,~K.; Nefedov,~A.;
  W\"{o}ll,~C.; Barth,~J.~V. {Self-Metalation of 2H-Tetraphenylporphyrin on
  Cu(111) An X-Ray Spectroscopy Study}. \emph{J. Chem. Phys.} \textbf{2012},
  \emph{136}, 014705\relax
\mciteBstWouldAddEndPuncttrue
\mciteSetBstMidEndSepPunct{\mcitedefaultmidpunct}
{\mcitedefaultendpunct}{\mcitedefaultseppunct}\relax
\EndOfBibitem
\bibitem[R\"{o}ckert et~al.(2014)R\"{o}ckert, Ditze, Stark, Xiao,
  Steinr\"{u}ck, Marbach, and Lytken]{Roeckert2014}
R\"{o}ckert,~M.; Ditze,~S.; Stark,~M.; Xiao,~J.; Steinr\"{u}ck,~H.-P.;
  Marbach,~H.; Lytken,~O. {Abrupt Coverage-Induced Enhancement of the
  Self-Metalation of Tetraphenylporphyrin with Cu(111)}. \emph{J. Phys. Chem.
  C} \textbf{2014}, \emph{118}, 1661--1667\relax
\mciteBstWouldAddEndPuncttrue
\mciteSetBstMidEndSepPunct{\mcitedefaultmidpunct}
{\mcitedefaultendpunct}{\mcitedefaultseppunct}\relax
\EndOfBibitem
\bibitem[Nowakowski et~al.(2013)Nowakowski, Wäckerlin, Girovsky, Siewert,
  Jung, and Ballav]{Nowakowski2013}
Nowakowski,~J.; W\"{a}ckerlin,~C.; Girovsky,~J.; Siewert,~D.; Jung,~T.~A.;
  Ballav,~N. {Porphyrin Metalation Providing an Example of a Redox Reaction
  Facilitated by a Surface Reconstruction}. \emph{Chem. Commun.} \textbf{2013},
  \emph{49}, 2347--2349\relax
\mciteBstWouldAddEndPuncttrue
\mciteSetBstMidEndSepPunct{\mcitedefaultmidpunct}
{\mcitedefaultendpunct}{\mcitedefaultseppunct}\relax
\EndOfBibitem
\bibitem[Goldoni et~al.(2012)Goldoni, Pignedoli, Santo, Castellarin-Cudia,
  Magnano, Bondino, Verdini, and Passerone]{Goldoni2012}
Goldoni,~A.; Pignedoli,~C.~A.; Santo,~G.~D.; Castellarin-Cudia,~C.;
  Magnano,~E.; Bondino,~F.; Verdini,~A.; Passerone,~D. {Room Temperature
  Metalation of 2H-TPP Monolayer on Iron and Nickel Surfaces by Picking up
  Substrate Metal Atoms}. \emph{ACS Nano} \textbf{2012}, \emph{6},
  10800--10807\relax
\mciteBstWouldAddEndPuncttrue
\mciteSetBstMidEndSepPunct{\mcitedefaultmidpunct}
{\mcitedefaultendpunct}{\mcitedefaultseppunct}\relax
\EndOfBibitem
\bibitem[Gupta et~al.(2007)Gupta, Bedi, and Mahajan]{Gupta2007}
Gupta,~H.; Bedi,~R.~K.; Mahajan,~A. {Characterization of Hot Wall Grown Silver
  Phthalocyanine Films}. \emph{J. Appl. Phys.} \textbf{2007}, \emph{102},
  073502\relax
\mciteBstWouldAddEndPuncttrue
\mciteSetBstMidEndSepPunct{\mcitedefaultmidpunct}
{\mcitedefaultendpunct}{\mcitedefaultseppunct}\relax
\EndOfBibitem
\bibitem[Karweik et~al.(1974)Karweik, Winograd, Davis, and Kadish]{Karweik1974}
Karweik,~D.; Winograd,~N.; Davis,~D.; Kadish,~K. {X-Ray Photoelectron
  Spectroscopic Studies of Silver(III) Octaethylporphyrin}. \emph{J. Am. Chem.
  Soc} \textbf{1974}, \emph{96}, 591--592\relax
\mciteBstWouldAddEndPuncttrue
\mciteSetBstMidEndSepPunct{\mcitedefaultmidpunct}
{\mcitedefaultendpunct}{\mcitedefaultseppunct}\relax
\EndOfBibitem
\bibitem[Karweik and Winograd(1976)Karweik, and Winograd]{Karweik1976}
Karweik,~D.; Winograd,~N. {Nitrogen Charge Distribution in Free-Base
  Porphyrins, Metalloporphyrins, and their Reduced Analogues Observed by X-Ray
  Photoelecton Spectroscopy}. \emph{Inorg. Chem.} \textbf{1976}, \emph{15},
  2336--2342\relax
\mciteBstWouldAddEndPuncttrue
\mciteSetBstMidEndSepPunct{\mcitedefaultmidpunct}
{\mcitedefaultendpunct}{\mcitedefaultseppunct}\relax
\EndOfBibitem
\bibitem[Fukuzumi et~al.(2008)Fukuzumi, Ohkubo, Zhu, Sintic, Khoury, Sintic, E,
  Ou, Crossley, and Kadish]{Fukuzumi2008}
Fukuzumi,~S.; Ohkubo,~K.; Zhu,~W.; Sintic,~M.; Khoury,~T.; Sintic,~P.~J.;
  E,~W.; Ou,~Z.; Crossley,~M.~J.; Kadish,~K.~M. {Androgynous Porphyrins.
  Silver(II) Quinoxalinoporphyrins Act as Both Good Electron Donors and
  Acceptors}. \emph{J. Am. Chem. Soc} \textbf{2008}, \emph{130},
  9451--9458\relax
\mciteBstWouldAddEndPuncttrue
\mciteSetBstMidEndSepPunct{\mcitedefaultmidpunct}
{\mcitedefaultendpunct}{\mcitedefaultseppunct}\relax
\EndOfBibitem
\bibitem[Sperl et~al.(2011)Sperl, Kr\"{o}ger, and Berndt]{Sper2011}
Sperl,~A.; Kr\"{o}ger,~J.; Berndt,~R. {Controlled Metalation of a Single
  Adsorbed Phthalocyanine}. \emph{Angew. Chem. Int. Ed.} \textbf{2011},
  \emph{50}, 5294--5297\relax
\mciteBstWouldAddEndPuncttrue
\mciteSetBstMidEndSepPunct{\mcitedefaultmidpunct}
{\mcitedefaultendpunct}{\mcitedefaultseppunct}\relax
\EndOfBibitem
\bibitem[Walch et~al.(2010)Walch, Gutzler, Sirtl, Eder, and
  Lackinger]{Walch2010}
Walch,~H.; Gutzler,~R.; Sirtl,~T.; Eder,~G.; Lackinger,~M. {Material- and
  Orientation-Dependent Reactivity for Heterogeneously Catalyzed Carbon-Bromine
  Bond Homolysis}. \emph{J. Phys. Chem. C} \textbf{2010}, \emph{114},
  12604--12609\relax
\mciteBstWouldAddEndPuncttrue
\mciteSetBstMidEndSepPunct{\mcitedefaultmidpunct}
{\mcitedefaultendpunct}{\mcitedefaultseppunct}\relax
\EndOfBibitem
\bibitem[Hanzl\'{\i}kov\'{a} et~al.(1998)Hanzl\'{\i}kov\'{a}, Proch\'{a}zka,
  \v{S}t\v{e}p\'{a}nek, Bok, Baumruk, and Anzenbacher]{Hanzlikova1998}
Hanzl\'{\i}kov\'{a},~J.; Proch\'{a}zka,~M.; \v{S}t\v{e}p\'{a}nek,~J.; Bok,~J.;
  Baumruk,~V.; Anzenbacher,~P. {Metalation of
  5,10,15,20-Tetrakis(1-methyl-4-pyridyl)porphyrin in Silver Colloids Studied
  via Time Dependence of Surface-Enhanced Resonance Raman Spectra}. \emph{J.
  Raman Spectrosc.} \textbf{1998}, \emph{29}, 575--584\relax
\mciteBstWouldAddEndPuncttrue
\mciteSetBstMidEndSepPunct{\mcitedefaultmidpunct}
{\mcitedefaultendpunct}{\mcitedefaultseppunct}\relax
\EndOfBibitem
\bibitem[\v{S}im\'{a}kov\'{a} et~al.(2014)\v{S}im\'{a}kov\'{a}, Gautier,
  Proch\'{a}zka, Herv\'{e}-Aubert, and Chourpa]{Simakova2014}
\v{S}im\'{a}kov\'{a},~P.; Gautier,~J.; Proch\'{a}zka,~M.; Herv\'{e}-Aubert,~K.;
  Chourpa,~I. {Polyethylene-glycol-Stabilized Ag Nanoparticles for
  Surface-Enhanced Raman Scattering Spectroscopy Ag Surface Accessibility
  Studied Using Metalation of Free-Base Porphyrins}. \emph{J. Phys. Chem. C}
  \textbf{2014}, \emph{118}, 7690--7697\relax
\mciteBstWouldAddEndPuncttrue
\mciteSetBstMidEndSepPunct{\mcitedefaultmidpunct}
{\mcitedefaultendpunct}{\mcitedefaultseppunct}\relax
\EndOfBibitem
\bibitem[Gorgoi and Zahn(2006)Gorgoi, and Zahn]{Gorgoi2006}
Gorgoi,~M.; Zahn,~D.~R. {Charge-Transfer at Silver Phthalocyanines Interfaces}.
  \emph{Appl. Surf. Sci.} \textbf{2006}, \emph{252}, 5453--5456\relax
\mciteBstWouldAddEndPuncttrue
\mciteSetBstMidEndSepPunct{\mcitedefaultmidpunct}
{\mcitedefaultendpunct}{\mcitedefaultseppunct}\relax
\EndOfBibitem
\bibitem[Seah and Dench(1979)Seah, and Dench]{Seah1979}
Seah,~M.~P.; Dench,~W.~A. {Quantitative Electron Spectroscopy of Surfaces: A
  Standard Data Base for Electron Inelastic Mean Free Paths in Solids}.
  \emph{Surf. Interface Anal.} \textbf{1979}, \emph{1}, 2--11\relax
\mciteBstWouldAddEndPuncttrue
\mciteSetBstMidEndSepPunct{\mcitedefaultmidpunct}
{\mcitedefaultendpunct}{\mcitedefaultseppunct}\relax
\EndOfBibitem
\bibitem[Horcas et~al.(2007)Horcas, Fernandez, Gomez-Rodriguez, Colchero,
  Gomez-Herrero, and Baro]{Horcas2007}
Horcas,~I.; Fernandez,~R.; Gomez-Rodriguez,~J.; Colchero,~J.;
  Gomez-Herrero,~J.; Baro,~A. {WSXM: A Software for Scanning Probe Microscopy
  and a Tool for Nanotechnology}. \emph{Rev. Sci. Instrum.} \textbf{2007},
  \emph{78}, 013705\relax
\mciteBstWouldAddEndPuncttrue
\mciteSetBstMidEndSepPunct{\mcitedefaultmidpunct}
{\mcitedefaultendpunct}{\mcitedefaultseppunct}\relax
\EndOfBibitem
\bibitem[Enkovaara et~al.(2010)Enkovaara, Rostgaard, Mortensen, Chen,
  Du{\l{}}ak, Ferrighi, Gavnholt, Glinsvad, Haikola, Hansen, Kristoffersen,
  Kuisma, Larsen, Lehtovaara, Ljungberg, Lopez-Acevedo, Moses, Ojanen, Olsen,
  Petzold, Romero, Stausholm-M{\o{}}ller, Strange, Tritsaris, Vanin, Walter,
  Hammer, Häkkinen, Madsen, Nieminen, N{\o{}}rskov, Puska, Rantala,
  Schi{\o{}}tz, Thygesen, and Jacobsen]{Enkovaara2010}
Enkovaara,~J. et~al.  {Electronic Structure Calculations with GPAW: a
  Real-Space Implementation of the Projector Augmented-Wave Method}. \emph{J.
  Phys.: Condens. Matter} \textbf{2010}, \emph{22}, 253202\relax
\mciteBstWouldAddEndPuncttrue
\mciteSetBstMidEndSepPunct{\mcitedefaultmidpunct}
{\mcitedefaultendpunct}{\mcitedefaultseppunct}\relax
\EndOfBibitem
\bibitem[Tkatchenko and Scheffler(2009)Tkatchenko, and
  Scheffler]{Tkatchenko2009}
Tkatchenko,~A.; Scheffler,~M. {Accurate Molecular van der Waals Interactions
  from Ground-State Electron Density and Free-Atom Reference Data}. \emph{Phys.
  Rev. Lett.} \textbf{2009}, \emph{102}, 073005\relax
\mciteBstWouldAddEndPuncttrue
\mciteSetBstMidEndSepPunct{\mcitedefaultmidpunct}
{\mcitedefaultendpunct}{\mcitedefaultseppunct}\relax
\EndOfBibitem
\bibitem[Larsen et~al.(2009)Larsen, Vanin, Mortensen, Thygesen, and
  Jacobsen]{Larsen2009}
Larsen,~A.~H.; Vanin,~M.; Mortensen,~J.~J.; Thygesen,~K.~S.; Jacobsen,~K.~W.
  {Localized Atomic Basis Set in the Projector Augmented Wave Method}.
  \emph{Phys. Rev. B: Condens. Matter Mater. Phys.} \textbf{2009}, \emph{80},
  195112\relax
\mciteBstWouldAddEndPuncttrue
\mciteSetBstMidEndSepPunct{\mcitedefaultmidpunct}
{\mcitedefaultendpunct}{\mcitedefaultseppunct}\relax
\EndOfBibitem
\bibitem[Morgenstern et~al.(1999)Morgenstern, L{\ae}gsgaard, Stensgaard, and
  Besenbacher]{Morgenstern1999}
Morgenstern,~K.; L{\ae}gsgaard,~E.; Stensgaard,~I.; Besenbacher,~F. {Transition
  from One-Dimensional to Two-Dimensional Island Decay on an Anisotropic
  Surface}. \emph{Phys. Rev. Lett.} \textbf{1999}, \emph{83}, 1613--1616\relax
\mciteBstWouldAddEndPuncttrue
\mciteSetBstMidEndSepPunct{\mcitedefaultmidpunct}
{\mcitedefaultendpunct}{\mcitedefaultseppunct}\relax
\EndOfBibitem
\bibitem[Smykalla et~al.(2014)Smykalla, Shukrynau, Mende, R\"{u}ffer, Lang, and
  Hietschold]{Smykalla2014a}
Smykalla,~L.; Shukrynau,~P.; Mende,~C.; R\"{u}ffer,~T.; Lang,~H.;
  Hietschold,~M. {Manipulation of the Electronic Structure by Reversible
  Dehydrogenation of Tetra(p-hydroxyphenyl)porphyrin Molecules}. \emph{Surf.
  Sci.} \textbf{2014}, \emph{628}, 92--97\relax
\mciteBstWouldAddEndPuncttrue
\mciteSetBstMidEndSepPunct{\mcitedefaultmidpunct}
{\mcitedefaultendpunct}{\mcitedefaultseppunct}\relax
\EndOfBibitem
\bibitem[Smykalla et~al.(2014)Smykalla, Shukrynau, Mende, R\"{u}ffer, Lang, and
  Hietschold]{Smykalla2014b}
Smykalla,~L.; Shukrynau,~P.; Mende,~C.; R\"{u}ffer,~T.; Lang,~H.;
  Hietschold,~M. {Interplay of Hydrogen Bonding and Molecule-Substrate
  Interaction in Self-Assembled Adlayer Structures of a
  Hydroxyphenyl-Substituted Porphyrin}. \emph{Surf. Sci.} \textbf{2014},
  \emph{628}, 132--140\relax
\mciteBstWouldAddEndPuncttrue
\mciteSetBstMidEndSepPunct{\mcitedefaultmidpunct}
{\mcitedefaultendpunct}{\mcitedefaultseppunct}\relax
\EndOfBibitem
\bibitem[Smykalla et~al.(2015)Smykalla, Shukrynau, Mende, Lang, Knupfer, and
  Hietschold]{Smykalla2015a}
Smykalla,~L.; Shukrynau,~P.; Mende,~C.; Lang,~H.; Knupfer,~M.; Hietschold,~M.
  {Photoelectron Spectroscopy Investigation of the Temperature-Induced
  Deprotonation and Substrate-Mediated Hydrogen Transfer in a
  Hydroxyphenyl-Substituted Porphyrin}. \emph{Chem. Phys.} \textbf{2015},
  \emph{450-451}, 39--45\relax
\mciteBstWouldAddEndPuncttrue
\mciteSetBstMidEndSepPunct{\mcitedefaultmidpunct}
{\mcitedefaultendpunct}{\mcitedefaultseppunct}\relax
\EndOfBibitem
\bibitem[Toader and Hietschold(2011)Toader, and Hietschold]{Toader2011a}
Toader,~M.; Hietschold,~M. {Tuning the Energy Level Alignment at the SnPc
  Ag(111) Interface Using an STM Tip}. \emph{J. Phys. Chem. C} \textbf{2011},
  \emph{115}, 3099--3105\relax
\mciteBstWouldAddEndPuncttrue
\mciteSetBstMidEndSepPunct{\mcitedefaultmidpunct}
{\mcitedefaultendpunct}{\mcitedefaultseppunct}\relax
\EndOfBibitem
\end{mcitethebibliography}

\end{document}